\documentclass[aps,prl,twocolumn,10pt,superscriptaddress]{revtex4-1}
\usepackage{float}
\usepackage{mathrsfs}
\usepackage{amsmath}
\usepackage{amssymb}
\usepackage{graphicx}
\usepackage{colordvi}
\newcommand*{\Hamiltonian}
{1}
\newcommand*{\dPdv}
{2}
\newcommand*{\VelScale}
{3}
\newcommand*{\Deltavsqrt}
{8}

\begin{document}

\title{Dissociation of one-dimensional matter-wave breathers due to quantum
many-body effects}
\author{Vladimir A. Yurovsky}
\affiliation{School of Chemistry, Tel Aviv University, 6997801 Tel Aviv,
Israel}
\email{volodia@post.tau.ac.il}
\author{Boris A. Malomed}
\affiliation{Department of Physical Electronics, School of Electrical Engineering,
Faculty of Engineering, Tel Aviv University, Tel Aviv 6997801, Israel} 
\affiliation{ITMO University, St. Petersburg 197101, Russia}
\author{Randall G. Hulet}
\affiliation{Department of Physics and Astronomy, Rice University, Houston,
TX 77005, USA}
\author{Maxim Olshanii}
\affiliation{Department of Physics, University of Massachusetts Boston,
Boston, MA 02125, USA}

\begin{abstract}
We use the ab initio  Bethe Ansatz dynamics to predict the dissociation of one-dimensional
cold-atom breathers that are created by a quench from a fundamental soliton.  
We find that the dissociation is a robust quantum 
many-body effect, while in the mean-field (MF) limit the 
dissociation is forbidden by the integrability of the underlying nonlinear
Schr\"{o}dinger equation. The analysis demonstrates the possibility to observe
quantum many-body effects without leaving the MF range of experimental
parameters. We find that the dissociation time is of the order of a few
seconds for a typical atomic-soliton setting.
\end{abstract}

\maketitle


Under normal conditions, interacting quantum Bose gases do not readily
exhibit signatures of their corpuscular nature, but rather follow the behavior
predicted by mean-field (MF) theory. The observability of microscopic
quantum effects involving a substantial fraction of the particles in a coherent
macroscopic setting generally requires going beyond-MF, for example, at low
density in 1D \cite{Kinoshita2004,tolra2004_190401} or high density in 3D.
In 3D systems, the high-density Lee-Huang-Yang corrections, which are induced by
quantum correlations, were realized experimentally using the Feshbach
resonance \cite{navon2011_135301} and in the spectacular form of
``quantum droplets" in dipolar 
\cite{ferrierbarbut2016_215301,schmitt2016,chomaz2016}\ and isotropic 
\cite{cabrera2017} bosonic gases., i.e.,\ as self-trapped states stabilized
against the collapse by the beyond-MF self-repulsion.  This stabilization was predicted
in Refs. \cite{petrov2015,petrov2016,baillie2016}. 
Quantum effects involving a macroscopic number of atoms in collapsing
attractive 3D gases and colliding condensates were also observed 
\cite{sackett1999,gerton2000,donley2001,vogels2002,chin2003} and analyzed 
\cite{yurovsky2002,yukalov2004} in the MF density range.

A generic opportunity to observe beyond-MF effects arises when a particular
symmetry of the MF dynamics, which prohibits a certain effect, is broken at
the microscopic level thus making observation of the effect possible. For
instance, the scale invariance in the dynamics of a harmonically trapped 2D
Bose gas nullifies the interaction-induced shift of the frequency of
monopole excitations for all excitation amplitudes; however, this scale
invariance is broken by the quantum many-body Hamiltonian, leading to a
small shift, albeit discernible on a zero background 
\cite{olshanii2010_095302}. In this context, the symmetry breaking by the
secondary quantization may be considered as a manifestation of a general
phenomenon known as the quantum anomaly \cite{gupta1993}. In this Letter we
develop a similar strategy for predicting beyond-MF effects in the
one-dimensional (1D) self-attractive Bose gas in a MF range of parameters.
The respective MF equation amounts to the nonlinear Schr\"{o}dinger (NLS)
equation, integrable by the inverse-scattering transform 
\cite{ablowitz1981}.  The NLS rigidly links the structure of a 
time-dependent solution to its
initial form, with many features of the latter rendered identifiable in the
former. In particular, a sudden increase of
the strength of the attractive coupling constant by a factor of $4$, i.e. 
an interaction \textit{quench},
converts a fundamental soliton into an exact superposition of two solitons
with zero relative velocity, zero spatial separation, and with a mass ratio $3:1$  
\cite{zakharov1972_62,satsuma1974_284,carr2002}. The two superimposed
solitons have different chemical potentials, hence the density oscillates as
a result of interference. Such an exact superposition of fundamental
solitons is identified as an NLS breather.

Further, quantum fluctuations in solitons have also been analyzed in terms
of the exact Bethe-ansatz (BA) solution 
\cite{berezin1964,mcguire1964_622,calogero1975,lai1989_854}, the linearization
approximation \cite{lai1989_844}, and the numerical positive-$P$ representation
\cite{carter1987,drummond1987}. These effects have been observed in
experiments \cite{rosenbluh1991,drummond1993,corney2006,corney2008}, see
also review \cite{drummond2016}. In particular, in the quantum many-body
theory, contrary to its MF counterpart, the center-of-mass (COM)
position of a soliton is a quantum coordinate whose conjugate momentum is
subject to quantum fluctuations 
\cite{drummond1994,vaughan2007,castin2009_317}.

The MF breather generated by the quench does not split due to the absence of 
any relative velocity in the MF.  We predict, however, that the spread
of the relative velocity of the two
solitons leads to dissociation, and thus reveals a many-body quantum effect.
A different dissociation scenario was predicted in \cite{streltsov2008}.

In Ref. \cite{weiss2016} it is shown, using a Bose-Hubbard model, that
higher-order solitons also break up due to many-body quantum effects. The
fact that a nonintegrable lattice model, with thermalization of eigenstates, also
predicts many-body quantum effects is relevant for comparison with
results of the present work.

We consider $N$ atoms of mass $m$ moving in a waveguide with a transverse
trapping frequency $\omega _{\perp }$. In the ``deep 1D" approximation they
can be considered as particles moving in the $x$ direction, with zero-range
interactions of strength $g=2\hbar a\omega _{\perp }$ \cite{olshanii1998_938},
where the $s$-wave scattering length $a$ can be tuned by an external
magnetic field via the Feshbach resonance. The corresponding Lieb-Liniger\
Hamiltonian is \cite{yurovsky2008b}
\begin{equation}
\hat{H}=-\frac{\hbar ^{2}}{2m}\sum_{j=1}^{N}\frac{\partial ^{2}}{\partial
x_{j}^{2}}+g\sum_{j<j^{\prime }}\delta (x_{j}-x_{j^{\prime }})\,\,.
\label{eq:Hamiltonian}
\end{equation}%
This problem has an exact BA solution \cite{lieb1963,berezin1964}. Due to
the translational invariance of the Hamiltonian (\ref{eq:Hamiltonian}), its
eigenfunctions are delocalized, having a homogeneous density. For attractive
interactions with $g<0$, there are also eigenstates in the form of one or
several several strings -- bound states of several particles, i.e., quantum
solitons \cite{berezin1964,mcguire1964_622,yang1968}. Although they remained
a theoretical concept since they were introduced, very recently similar
states --- the Bethe strings --- have been directly observed in an
antiferromagnetic Heisenberg-Ising chain \cite{wang2017}). A superposition
of strings with different velocities may remain localized for a finite time,
so that it carries over into a MF multi-soliton (breather) in the limit of 
$N\rightarrow \infty $ \cite{calogero1975,lai1989_844,lai1989_854}.
Normalization factors for multi-string states were derived in
Ref. \cite{calabrese2007l}.

We assume that, at $t<0$, the interaction strength was $g_{0}=g/4$, and the
system contained a single-string state $\varphi _{N}^{(0)}$ with zero COM
velocity. At $t=0$, the external magnetic field suddenly changes, switching
the interaction strength to $g$, i.e., applying a $4$-fold \textit{quench}
to the system. The exact BA calculation, starting from the quenched state,
makes it possible to directly compare the result in the quantum many-body
system with its exactly known MF counterpart --- the second-order breather,
which is generated by the $4$-fold quench \cite{satsuma1974_284}. This is,
essentially, the objective of the present work. 

\begin{figure}
\includegraphics[width=3.4in]{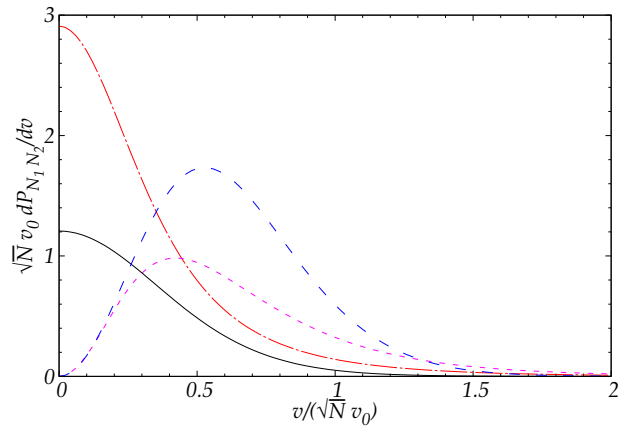}
\caption{Channel-selective probability distributions for the relative
velocity [see Eq. (\protect\ref{eq:dPdv})] of the dissociation products,
produced by the application of the quench to the single string (fundamental
quantum soliton). The black solid and red dot-dashed lines show 
$dP_{15,5}/dv $ and $dP_{3,1}/dv$, for $N=20$ and $4$, respectively and the
same ratio, $N_{1}/\left( N-N_{1}\right) =3:1$, as the MF breather. Plots
for other ratios have similar shapes except for ones with $N_1=N/2$, which
are shown by the blue long and magenta short dashes ($10^{6}dP_{10,10}/dv$
and $10dP_{2,2}/dv$, respectively). The velocity scale $v_{0}$ is defined by
Eq. (\protect\ref{eq:VelScale}). }
\label{fig:dPdv}
\end{figure}

After the application of the quench, the many-body configuration will be a
superposition of a single-string state $\varphi _{N}$, double-string states 
$\varphi _{N_{1},N-N_1,v}$, where $v$ is the relative velocity of two strings
composed of $N_{1}$ and $N-N_{1}$ atoms, and multi-string states. On the
other hand, a fundamental quantum soliton is a superposition of the
single-string states with different COM velocities. These states are
mutually orthogonal due to the COM velocity conservation, therefore
probabilities of quench-triggered transitions from the pre-quench
fundamental-soliton state to multi-soliton ones will be the same as for the
delocalized string states. The probabilities are calculated analytically
using the exact BA solution \cite{supplement}. It is the basic technical
result of the present work, which underlies the physical considerations.
First, the probability to remain in the single-string state is
\[
\left\vert \left\langle \varphi _{N}^{(0)}|\varphi _{N}\right\rangle
\right\vert ^{2}=\left( \frac{2\sqrt{|gg_{0}|}}{|g|+|g_{0}|}\right)
^{2(N-1)}=\left( \frac{4}{5}\right) ^{2(N-1)}
\]
For the double-string states the probabilities depend on the relative string
velocity $v>0$ and the string composition,
\begin{equation}
\frac{dP_{N_{1},N-N_{1}}(v)}{dv}=(2-\delta _{N_{1},N-N_{1}})\left\vert
\left\langle \varphi _{N}^{(0)}|\varphi _{N_{1},N-N_{1},v}\right\rangle
\right\vert ^{2}\,\,.  \label{eq:dPdv}
\end{equation}%
It is a sum of the probabilities corresponding to velocities $v$ and $-v$
for $N_{1}\neq N-N_{1}$, while for $N_{1}=N/2$ the states with $v$ and $-v$
are identical. Examples of the probabilities are displayed in Fig. 
\ref{fig:dPdv}, for $N=4$ and $N=20$. The natural velocity scale is
\begin{equation}
v_{0}=|g|/(2\hbar )\equiv a\omega _{\perp }  \label{eq:VelScale}
\end{equation}

\begin{figure}
\includegraphics[width=3.4in]{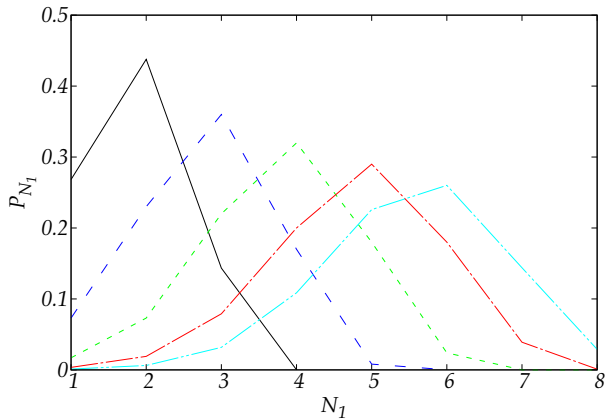}
\caption{Total probabilities for different dissociation channels (\protect
\ref{eq:PN1}), produced by the application of the $g/4\rightarrow g$ quench
to the single string (fundamental quantum soliton) composed of 
$N=8,\,12,\,16,\,20$, and $23$ atoms (black solid, blue long-dashed, green
short-dashed, red dot-dashed, and cyan dot-dot-dashed lines, respectively). }
\label{fig:ProbN1}
\end{figure}

Total probabilities of the transition to double-string states with fixed
$N_{1}$,
\begin{equation}
P_{N_{1}}\equiv \int_{0}^{\infty }\frac{dP_{N_{1},N-N_{1}}(v)}{dv}dv,
\label{eq:PN1}
\end{equation}%
are presented in Fig. \ref{fig:ProbN1} making it obvious that the transition
$N\rightarrow 3N/4+N/4$ features the \emph{largest probability}, in
agreement with the MF prediction. The cumulative probability of the
transition to all double-string states, 
$\sum_{N_{1}=1}^{\left[ N/2\right]}P_{N_{1}}$, exceeds $80\%$ for $N\geq 8$ 
(here, $\left[ ...\right]$ stand for the integer part).

Another similarity to the MF is seen in the fact that the quench-produced
configuration, being a superposition of multi-string eigenstates with
different energies, oscillates in time due to their interference, thus
qualitatively resembling the breather. The binding energy of the
multi-string solution is the sum of the constituting string energies, each
one being $E_{N_{1}}=-N_{1}(N_{1}^{2}-1)mg^{2}/(24\hbar ^{2})$ 
\cite{berezin1964,mcguire1964_622}. In particular, the binding-energy difference
between the $(N_{1},N-N_{1})$ and $(N_{1}-1,N-N_{1}+1)$ double-string states
leads to beatings at frequency 
$[E_{N_{1}-1}+E_{N-N_{1}+1}-(E_{N_{1}}+E_{N-N_{1}})]/\hbar=N[N_{1}-(N+1)/2]mg^{2}/(4\hbar ^{3})$, 
which tends to the the MF
breather frequency, $mg^{2}N^{2}/(16\hbar ^{3})$, at 
$N_{1}=3N/4\rightarrow\infty$.

\begin{figure}
\includegraphics[width=3.4in]{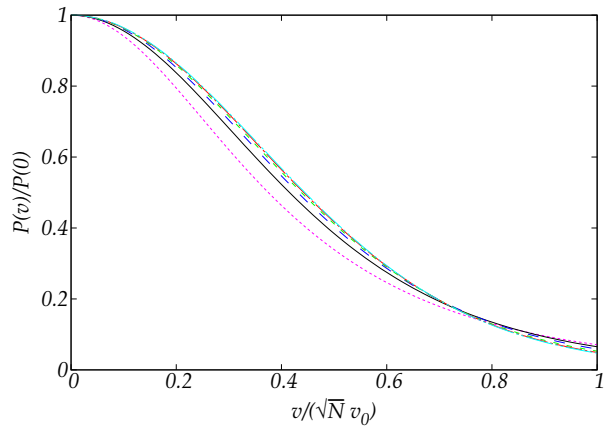}
\caption{Probability distributions [see Eq. (\protect\ref{eq:probv})],
totalled over all double-string dissociation channels, for the relative
velocity of the emerging strings, as produced by the application of the
quench to the single string (fundamental quantum soliton) composed of 
$N=4,\,8,\,12,\,16,\,20$, and $23$ particles (magenta dotted, black solid,
blue long-dashed, green short-dashed, red dot-dashed, and cyan
dot-dot-dashed lines, respectively, the last three lines being almost
indistinguishable). Velocity scale $v_{0}$ is taken as per 
Eq. (\protect\ref{eq:VelScale}). }
\label{fig:probv}
\end{figure}

Probability distributions for the relative velocity of the dissociation
products, summed up over all double-string dissociation channels,
\begin{equation}
P(v)=\sum_{N_{1}=1}^{\left[ N/2\right] }\frac{dP_{N_{1},N-N_{1}}(v)}{dv},
\label{eq:probv}
\end{equation}%
is almost independent of $N$, see its plot as a function of $v/\sqrt{N}$ in
Fig. \ref{fig:probv}.

\begin{figure}
\includegraphics[width=3.4in]{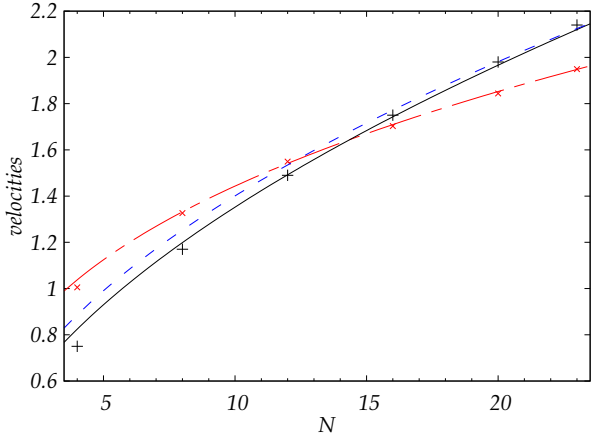}
\caption{HWHM (pluses) and r.m.s.\ (crosses) values of the relative velocity
averaged over all double-string (two-soliton) dissociation channels, as a
function of the number of atoms, $N$. Fits provided by 
Eqs. (\protect\ref{Delta_v_fit}), (\protect\ref{RMSv}), and 
(\protect\ref{Delta_v_sqrt}) are
shown by the black solid, red dot-dashed, and blue dashed lines,
respectively. The velocity unit is $v_{0}$ 
[see Eq. (\protect\ref{eq:VelScale})].}
\label{fig:Vaver}
\end{figure}

The numerically calculated half-width at half-maximum (HWHM), $\Delta v$, of
the velocity distribution, defined by $P(\Delta v)=P(0)/2$, can be fitted to
the following formula, which is, naturally, close to the $\sqrt{N}$
dependence:
\begin{equation}
\Delta v\approx 0.39N^{0.54}v_{0},  \label{Delta_v_fit}
\end{equation}%
see Fig. \ref{fig:Vaver}. The relative velocity can be measured also by its
mean-square value,
\[
\left\langle v^{2}\right\rangle =\int_{0}^{\infty
}v^{2}P(v)dv/\int_{0}^{\infty }P(v)dv
\]
However, the numerically found root-mean-square (r.m.s.) velocity increases
with $N$ only as
\begin{equation}
\sqrt{\langle v^{2}\rangle }\approx 0.63N^{0.36}v_{0},  \label{RMSv}
\end{equation}%
according to the fit displayed in Fig. \ref{fig:Vaver}. The probability
distribution (\ref{eq:dPdv}) has slowly decaying tails for small $N$, in
particular, $dP_{3N/4,N/4}(v)/dv\sim v^{-3N}$ at $v\rightarrow \infty $. The
tails increase the r.m.s. velocity at small $N$ and, therefore, slower its
gain with $N$. On the contrary, due to the normalization condition, the
tails exhaust the width of the central part of the $v$ distribution at small 
$N$, boosting the HWHM growth with $N$. Then at large $N$, when the tail
effects fade out, the r.m.s. velocity and HWHM should gain faster and
slower, respectively, than at small $N$. These arguments suggest that both
measures of the relative velocity variation assume the same asymptotic
scaling at large $N$, which should be close to $\sqrt{N}$, according to 
Fig. \ref{fig:probv}. The eventual fit is displayed in Fig. \ref{fig:Vaver}:
\begin{equation}
\Delta v\approx 0.44\sqrt{N}v_{0}.  \label{Delta_v_sqrt}
\end{equation}

The following estimate confirms the $\sqrt{N}$ scaling for a typical
relative velocity of the solitons, $\delta v$. Consider the system placed in
an external harmonic-oscillator (HO) potential with frequency $\Omega $.
Varying $\Omega $ from vanishingly small values towards very large ones, at
each $\Omega $ one can apply the $g/4\rightarrow g$ quench to the respective
ground state. The figure of merit to monitor is $\delta \tilde{x}$---the
time-averaged distance, further symmetrized over permutations, between COMs
of two groups of atoms, each containing the number of atoms $\sim N$. At
small $\Omega $, the state obtained right after the quench is unaffected by
the external confinement, hence the two solitons (strings) start their
motion with the free-space relative velocity $\delta v$. Thus, the distance 
$\delta \tilde{x}$ will be dominated by the typical distance between the
solitons placed in the HO potential, $\delta v/\Omega $, which diverges at
small $\Omega $. This very long scale governs the estimate for 
$\delta\tilde{x}$, the other potentially relevant length scale, the average
distance between two atoms inside the same soliton, which is on the order of
the size of an individual soliton, $\sim \hbar^{2}/(m|g|N)$, does not diverge
at $\Omega \rightarrow 0$. Thus,
\[
\delta \tilde{x}_{\Omega \rightarrow 0}\sim \delta v/\Omega \,\,.
\]
On the other hand, at large $\Omega $, the effect of the interatomic
interactions vanishes and the estimate for $\delta \tilde{x}$ is determined
by zero-point quantum fluctuations of the COM position of the cloud
containing $\sim N$ particles:
\[
\delta \tilde{x}_{\Omega \rightarrow \infty } \sim 
\sqrt{\hbar /(Nm\Omega )}\,\,.
\]
A crossover between the two regimes occurs when the interaction energy per
particle (comparable to the chemical potential of the gas, $\mu $),
estimated as $\sim \mu \sim mg^{2}N^{2}/\hbar ^{2}$, becomes comparable to
the HO quantum, $\hbar \Omega $. Indeed, when the former is dominated over
by the latter, the interactions are irrelevant, and the system becomes an
HO-confined ideal gas. At the crossover, the two above-mentioned estimates
yield the same value. An estimate for $\delta v$ immediately follows:
\begin{gather*}
\delta \tilde{x}_{\Omega \rightarrow 0}|_{\mu \sim \hbar \Omega }\sim \delta
\tilde{x}_{\Omega \rightarrow \infty }|_{\mu \sim \hbar \Omega
}\,\,\Rightarrow \,\, \\
\qquad \delta v\sim \sqrt{\frac{\hbar \Omega }{Nm}}\Big|_{\Omega \sim 
\frac{mg^{2}N^{2}}{\hbar ^{3}}}\sim \frac{|g|}{\hbar }\sqrt{N}\,\,.
\end{gather*}%
Indeed, this estimate is consistent with the fit (\ref{Delta_v_sqrt}).


The above results suggest that experimental observation of the variance in
the relative velocity of the solitons due to quantum many-body effects may
be possible. To demonstrate this, we consider $3\times 10^{3}$ $^{7}$Li
atoms, in a waveguide with transverse trapping frequency 
$\omega _{\perp}=2\pi \times 254$ Hz. The initial state is a fundamental matter-wave
soliton, existing at scattering length $a_{t<0}=-1.0\,a_{\text{Bohr}}$,
which is quenched up to $a_{t>0}=-4\,a_{\text{Bohr}}$ 
\footnote{For the present set of parameters, the number of atoms is only two times
lower than the collapse critical number, $N_{\text{crit}}=7900$, hence
deviations from the purely 1D behavior are expected. However, moderate 3D
corrections do not destroy MF breathers \cite{golde2017}}. The resulting
state constitutes an NLS breather with an aphelion density profile
proportional to $\text{sech}^{2}(x/\ell _{\text{breather}})$ and width 
$\ell_{\text{breather}}=8\hbar ^{2}/(mgN)=36$ $\mathrm{\mu }$m 
\cite{zakharov1972_62,satsuma1974_284}. Assuming that the splitting of the
breather into two solitons becomes apparent when the distance between their
COMs, after evolution time $\tau $, $\Delta x=\Delta v\cdot \tau $, becomes
comparable to the breather's width $\ell _{\text{breather}}$, and using
extrapolation (\ref{Delta_v_fit}) for the relative velocity of the solitons,
we obtain $\tau \simeq 3$ s for the time necessary to certainly observe the
splitting of the breather caused by the quantum dynamics.

The predicted dissociation time can be made even shorter at the expense of
reducing the cloud population, assuming that the scattering length
simultaneously increases so as to keep product $Na$ at a finite fraction of
the collapse critical value, 
$Na\lesssim a_{\perp }$, $a_{\perp }\sim \sqrt{\hbar /(m\omega _{\perp })}$ 
being the size of the transverse vibrational
ground state of the waveguide used. The microscopic velocity scale $v_{0}$,
the separation velocity $\Delta v$, and the breather size 
$\ell _{\text{breather}}$ can be estimated as 
$v_{0}\lesssim \hbar /(ma_{\perp }N)$, 
$\Delta v\lesssim \hbar /(ma_{\perp }\sqrt{N})$, and 
$\ell _{\text{breather}}\gtrsim a_{\perp }$, respectively. Then the breather 
dissociation time diminishes as 
$\tau \sim \ell _{\text{breather}}/\Delta v\gtrsim (1/\omega_{\perp })\sqrt{N}$ 
with the decrease of the number of particles.

For the analysis of possibilities for the experimental implementation of the
predictions reported here, it is important to estimate deviations of
real-world settings from the idealized model 
\cite{Kivshar1989_763,Sakaguchi2004_066613,Yanay2009_033145}. In this connection,
it is essential to consider the departure from the one-dimensionality, as
suggested, in particular, by the work aimed at experimental observation of
the quantum violation of the scale-invariance-induced constancy of the
monopole-mode frequency in the 2D Bose gas. In that case, weak dependence of
the quantum state on the third, confined dimension tends to mask the quantum
many-body effects \cite{merloti2013_033007}. Nevertheless, experiments have
clearly demonstrated that 3D experimental setups with appropriately designed
transverse confinement can be efficiently used for the emulation of ideal
one-dimensional quantum settings, and such emulations are stable against
real-world disturbances. Relevant examples are the creation of the atomic
Newton's cradle with repulsive interactions \cite{kinoshita2006}, and the
realization of the super-Tonks-Girardeau gas \cite{haller2009}. The latter
example is especially relevant for the comparison with the present analysis,
as it is also based on attractive interactions. Predictions of the MF
counterpart of the Lieb-Liniger model, i.e., the Gross-Pitaevskii equation,
which are also based on the one-dimensionality and integrability, are very
well confirmed in numerous experiments with matter-wave solitons 
\cite{strecker2002,khaykovich2002,marchant2013,nguyen2014,marchant2016}. The
well-known stability of the exact solution of the Lieb-Liniger model 
\cite{lieb1963,berezin1964} clearly means that the results may only be slightly
perturbed by other distortions, such as external fluctuations and
inhomogeneities.

As concerns the full 3D analysis, an example which makes it possible to
explicitly compare the MF approximation and its many-body counterpart is
offered by the problem of the stabilization of the gas of bosons with
repulsive interactions, attracted to the center with potential $\sim -r^{-2}$. 
In that case, the MF predicts suppression of the quantum collapse and
creation of a ground state which is missing in the single-particle
formulation \cite{sakaguchi2011}, while the full many-body analysis
demonstrates that the same newly created state exists as a metastable one
\cite{astrakharchi2015}.

To summarize, we have showed that the dissociation of the 1D matter-wave
breather, initiated by the quench from the fundamental soliton, is a purely
quantum many-body effect, as all the MF contributions to the dissociation
vanish due to the integrability at the MF level. This conclusion opens the
way to observe truly quantum many-body effects without leaving the MF range
of experimental parameters. We have evaluated the dissociation time
corresponding to typical experimental parameters for atomic solitons. The
extrapolation of the present results to a larger number of atoms is
justified \cite{supplement} by the comparison with recent results produced
by truncated Wigner calculations in Ref. \cite{opanchuk2017}.
Both \cite{opanchuk2017} and our work predict a single gradually-expanding 
cloud, unlike the abrupt formation of two flying apart fragments, predicted 
in the previous work \cite{streltsov2008}.

We acknowledge financial support provided jointly by the National Science
Foundation, through grants PHY-1402249, PHY-1408309, PHY-1607215, and
PHY-1607221, and Binational (US-Israel) Science Foundation through grant
2015616, as well as support from the Welch Foundation (grant C-1133), the
Army Research Office Multidisciplinary University Research Initiative (grant
W911NF-14-1-0003), the Office of Naval Research, and Israel Science Foundation 
(grant 1287/17). We thank L. D. Carr, 
P. Drummond, V. Dunjko, and C. Weiss for valuable discussions.

\clearpage

\renewcommand{\theequation}{S-\arabic{equation}}
\renewcommand{\thefigure}{S\arabic{figure}}
\setcounter{equation}{0}
\setcounter{figure}{0}
\begin{widetext}
\begin{center}
{\large \bf Supplemental material for: Dissociation of one-dimensional matter-wave breathers due to quantum
many-body effects}\\
Vladimir A. Yurovsky, 
 Boris A. Malomed,  
 Randall G. Hulet, 
 and Maxim Olshanii
\end{center}
\end{widetext}
\section{Calculation of overlaps}

The present analysis is based on the exact quantum Bethe-ansatz
eigenfunctions \cite{lieb1963,berezin1964,yurovsky2008b} of the Hamiltonian 
(\Hamiltonian) in the main text, 
\begin{equation}
\left\vert \varphi \right\rangle =\mathscr{\mathcal{N}}
\sum_{\mathcal{P}}A(\mathcal{P})
\exp \left( i\sum_{j=1}^{N}p_{\mathcal{P}j}x_{j}/\hbar \right) ,
\label{eq:Bethe}
\end{equation}%
which are defined in the simplex $-L/2\leq x_{1}\leq x_{2}\leq \cdots \leq
x_{N}\leq L/2$ with periodic boundary conditions in the box 
$\left[ -L/2,L/2\right] $. The coefficients \cite{lieb1963,yurovsky2008b} 
\begin{equation}
A(\mathcal{P})=\prod_{j<j'}
\left[ 1-\frac{img}{\hbar(p_{\mathcal{P}j'}-p_{\mathcal{P}j})}\right]   
\label{eq:coeffA}
\end{equation}%
are expressed in terms of rapidities $p_{j}$. Solution (\ref{eq:Bethe}) can
describe several bound states (strings) 
\cite{berezin1964,mcguire1964_622,yang1968,calabrese2007l} of $N_{i}$ atoms with
center-of-mass momenta $P_{i}$. Coefficients $A(\mathcal{P})$ vanish if the
permutations $\mathcal{P}$ change ordering within a string. This provides
exponential decay of the wavefunction (\ref{eq:Bethe}) when the distance
between atoms tends to infinity. For multi-string solutions the
normalization factor $\mathcal{N}$ is derived in Ref. \cite{calabrese2007l}.

In the center-of-mass system, the single-string state 
$\left\vert \varphi_{N}\right\rangle $ 
\cite{berezin1964,mcguire1964_622} has rapidities 
\begin{equation}
p_{j}=i\frac{mg}{2\hbar }(N-2j+1)  \label{eq:rap1str}
\end{equation}%
and the only non-vanishing coefficient being $A(\mathcal{E})=N!$, where 
$\mathcal{E}$ is the identity permutation. The normalization factor 
\begin{equation}
\mathcal{N}_{N}=\frac{1}{N\sqrt{L}}\left( \frac{m|g|}{\hbar ^{2}}\right)
^{(N-1)/2}  \label{eq:norm1str}
\end{equation}%
provides $\left\langle \varphi _{N}|\varphi _{N}\right\rangle =1$.

In the case of the double-string state $\left\vert \varphi
_{N_{1}N_{2}v}\right\rangle $, containing $N_{1}$ and $N_{2}=N-N_{1}$ atoms,
respectively, in each string, rapidities $p_{j}$ are given by

\begin{equation}
p_{j}=%
\begin{cases}
P_{1}+i\frac{mg}{2\hbar }(N_{1}-2j+1), & 1\leq j\leq N_{1}, \\ 
P_{2}+i\frac{mg}{2\hbar }(N+N_{1}-2j+1), & N_{1}<j\leq N~.%
\end{cases}
\label{eq:rap2str}
\end{equation}%
Here the string center-of-mass momenta $P_{1}=mvN_{2}/N$ and 
$P_{2}=-mvN_{1}/N$ are related to the relative string velocity $v$ . 
The normalization factor,%
\begin{gather*}
\mathcal{N}_{N_{1}N_{2}v}=\left( \frac{m}{2\pi NN_{1}N_{2}\hbar L}\right)
^{1/2}\left( \frac{m|g|}{\hbar ^{2}}\right) ^{(N-2)/2} \\
\times \left( \frac{v^{2}+(N_{1}-N_{2})^{2}v_{0}^{2}}
{v^{2}+N^{2}v_{0}^{2}}\right) ^{1/2},
\end{gather*}%
is calculated using results from Ref. \cite{calabrese2007l}. It provides
$\left\langle \varphi _{N_{1}N_{2}v^{\prime }}|\varphi
_{N_{1}N_{2}v}\right\rangle \rightarrow \delta (v-v^{\prime })$ in the limit
of $L\rightarrow \infty $. The velocity scale $v_{0}$ is defined by
Eq. (\VelScale) in the main text.

Rapidities of the initial state $\left\vert \varphi _{N}^{(0)}\right\rangle $
are given by Eq. (\ref{eq:rap1str}) where $g$ is replaced by $g_{0}=g/4$.
The overlap 
\begin{gather}
\left\langle \varphi _{N}^{(0)}|\varphi _{N_{1}N_{2}v}\right\rangle =
\mathcal{N}_{N}^{(0)}\mathcal{N}_{N_{1}N_{2}v}N!\hbar ^{N-1}L
\sum_{\mathcal{P}}A(\mathcal{P}) 
\nonumber
\\
\times \prod_{l=1}^{N-1}\left\{ \sum_{j=1}^{l}\left[ ip_{\mathcal{P}j}-
\frac{mg_{0}}{2\hbar }(N-2j+1)\right] \right\} ^{-1},  
\label{eq:overlap}
\end{gather}%
calculated using Eq. (\ref{eq:Bethe}), is independent of $L$. Here
rapidities $p_{j}$ are given by Eq. (\ref{eq:rap2str}) and normalization
factor $\mathcal{N}_{N}^{(0)}$ is given by Eq. (\ref{eq:norm1str}) with $g$
replaced by $g_{0}$. This overlap is used in Eq. (\dPdv) in the main text.

Whenever $v=0$ we have $P_{1}=P_{2}$ and Eq. (\ref{eq:rap2str}) gives 
$p_{j}=p_{j+N/2}$. This means that if $N$ is an even number, i.e. $N_{1}$ and 
$N_{2}$ have the same parities, the set (\ref{eq:rap2str}) contains equal
rapidities. Therefore, coefficients $A(\mathcal{P})$ contain divergent
factors at $v=0$. In addition, divergent factors may appear in 
Eq. (\ref{eq:overlap}) whenever the sum in the curly brackets vanishes. Divergences
are canceled if all terms in Eq. (\ref{eq:overlap}) are combined with a
common denominator. Since the number of terms in the sum over permutations 
$\mathcal{P}$ between strings may be as large as $\sim 2^{N}$, this transformation
was performed using the computer algebra system Maxima \cite{maxima}. Nevertheless, the
system runs out of available computer resources when $N$ exceeds $20$.

For odd $N$ Eq. (\ref{eq:overlap}) does not contain divergences. However,
due to the loss of significance, the calculations become unreliable for
$N>23$.

\section{Approximate density evolution}

The mean-field Gross-Pitaevskii equation has an exact solution corresponding
to two scattering solitons \cite{gordon1983}. We are interested in the case
when the solitons contain $N_{1}=3N/4$ and $N_{2}=N/4$ atoms and their
relative phase and the distance between them are equal to zero at $t=0$,
such that the solution coincides with the breather at $t=0$. The mean-field
value at the center of mass ($x=0$) is expressed in terms of the scaled time 
$\tilde{t}=2\pi t/T_{\mathrm{breather}}$ and velocity $\tilde{v}%
=2v/(Nv_{0})$,%
\begin{widetext}
\[
u(0,\tilde{t})=\exp\left(i\frac{5}{8}(1-\tilde{v}^{2})\tilde{t}\right)
\frac{(1+\tilde{v}^{2})\cosh z\cos\phi+2i(\tilde{v}^{2}+1/4)\cosh z\sin\phi+3/2\tilde{v}\sinh z\sin\phi}{1/4+\tilde{v}^{2}+(1+\tilde{v}^{2})\cosh2z+3/4\cos2\phi},
\]
\end{widetext}
where $z=-3\tilde{v}\tilde{t}/4$, 
$\phi =(1+\tilde{v}^{2})\tilde{t}/2$, and 
$T_{\mathrm{breather}}=32\pi \hbar ^{3}/(mg^{2}N^{2})$ is
the classical breather period.

Let us approximate the system density by the two-soliton density averaged
over their relative velocity $v$, 
\begin{equation}
\rho (x,t)=\mathrm{const}\intop_{-\infty }^{\infty }dv
\frac{dP_{N_{1}N_{2}}(v)}{dv}|u(x,\tilde{t})|^{2}.  \label{eq:rhoxt}
\end{equation}%
The probability distribution can be approximated by a Gaussian, 
\begin{equation*}
\frac{dP_{N_{1}N_{2}}(v)}{dv}\approx \mathrm{const}
\exp \left( -\frac{d_{v}}{N}\left( \frac{v}{v_{0}}\right) ^{2}\right) ,
\end{equation*}%
where $d_{v}=(\ln 2)/0.44^{2}\approx  3.58$
provides the fit 
\begin{equation}
\Delta v\approx 0.44\sqrt{N}v_{0}  \label{eq:Delta_v_sqrt}
\end{equation}%
{[}see Eq. (\Deltavsqrt) in the main text{]}.

\begin{figure}[H]
\includegraphics[width=3.4in]{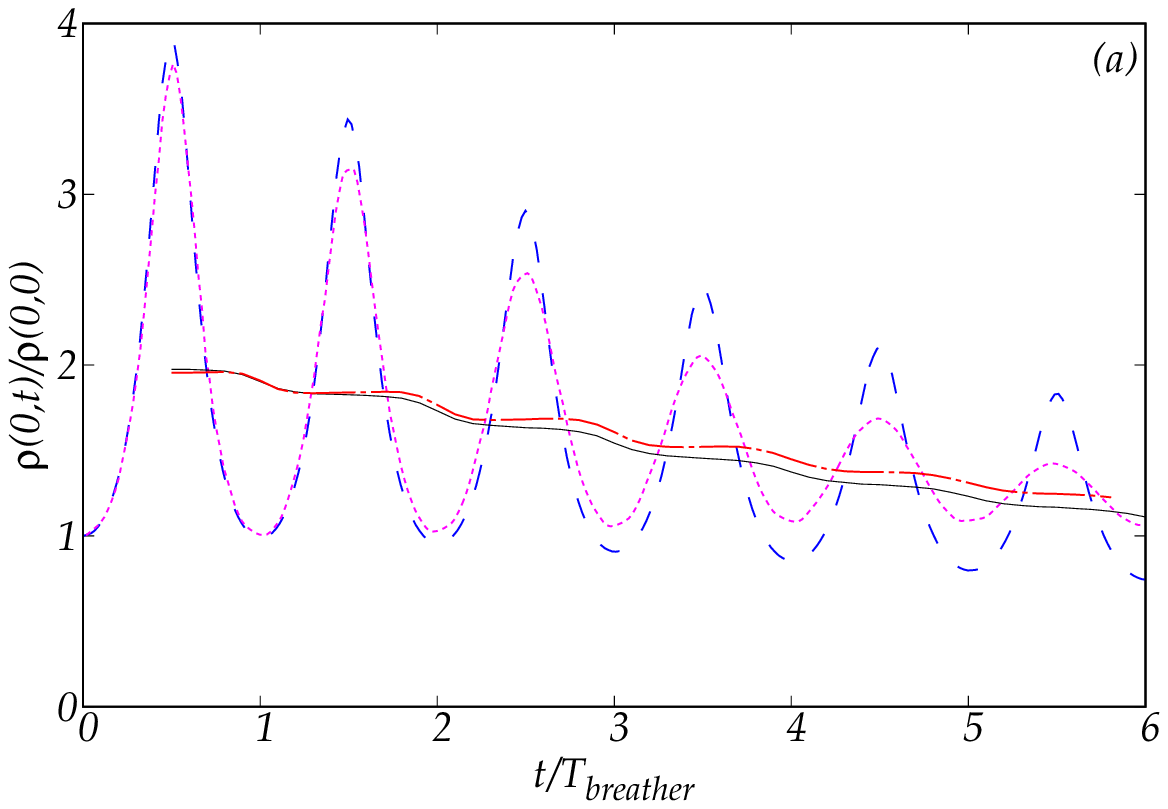}
\par
\includegraphics[width=3.4in]{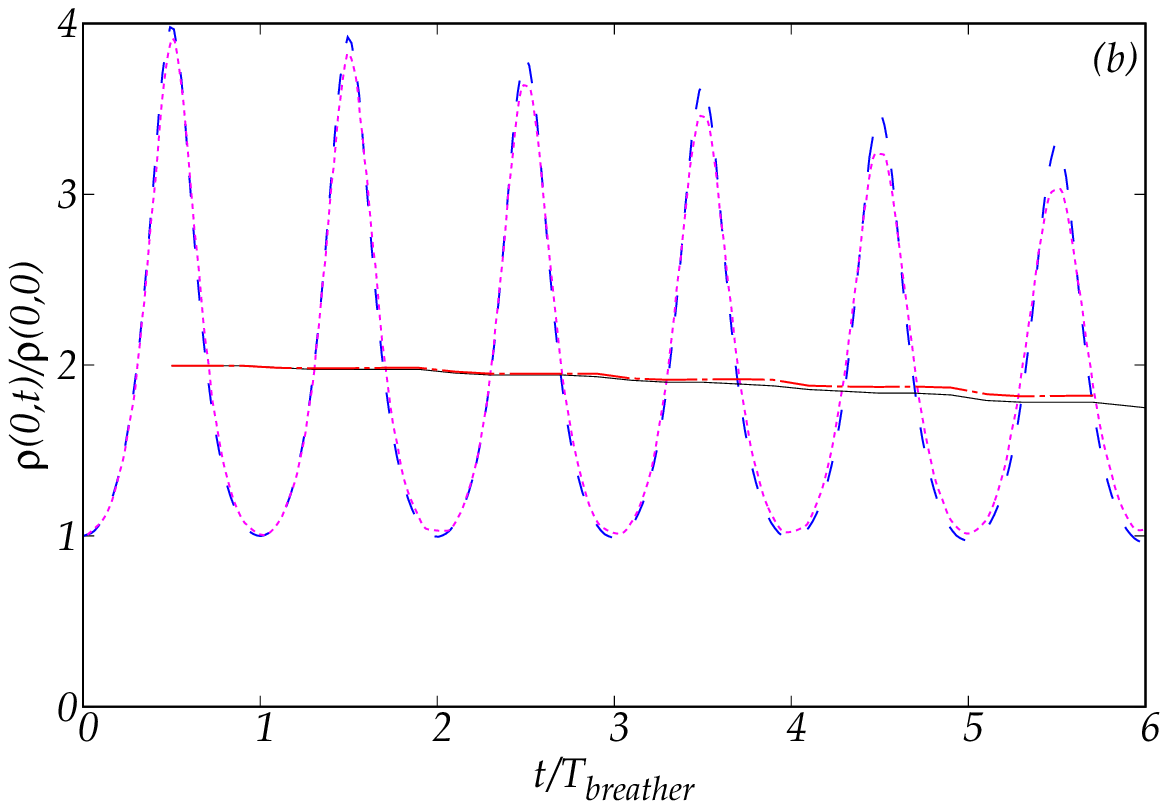}
\caption{Two-soliton density (\protect\ref{eq:rhoxt}), averaged over the
relative velocity, shown by the blue dashed line, is compared to results
produced by the truncated Wigner calculations in 
Ref. \protect\cite{opanchuk2017} (the magenta dotted line) 
for (a) $N=1000$ and (b) $N=10000$.
The values averaged over the breather period are shown by the
black solid and red dot-dashed lines, respectively.}
\label{fig:rho0t}
\end{figure}

The averaged mean-field density $\rho (0,t)$, calculated by numerical integration over $v$ in 
Eq. (\ref{eq:rhoxt}), is compared in Fig. \ref{fig:rho0t} to results produced by
truncated Wigner calculations in Ref. \cite{opanchuk2017}. The 
averaged mean-field neglects interference between states with
different relative velocities and correlations during evolution. This leads
to the underestimated damping of the breather oscillation amplitude. However, the
 average (over the breather period) values, which decrease due to the breather's
dissociation, are in a good agreement. This justifies the extrapolation of the
scaling low (\ref{eq:Delta_v_sqrt}) to large $N$.

\end{document}